\documentclass[graybox]{svmult}

\usepackage{mathptmx}       
\usepackage{helvet}         
\usepackage{courier}        
\usepackage{type1cm}        
\usepackage{makeidx}         
\usepackage{graphicx}
\usepackage{bm}
\usepackage{multicol}        
\usepackage[bottom]{footmisc}

\makeindex             

\begin{document}

\title*{Dynamics of probabilistic labor markets: statistical physics perspective}
\titlerunning{Dynamics of probabilistic labor markets} 
\author{He Chen and Jun-ichi Inoue}
\authorrunning{H. Chen and J. Inoue}
\institute{He Chen and Jun-ichi Inoue \at Hokkaido University \email{jinoue@cb4.so-net.ne.jp}}
%
%
\maketitle

\abstract{
We introduce a toy probabilistic model to analyze job-matching processes in recent Japanese labor markets 
for university graduates by means of statistical physics. We show that the aggregation probability of each company is 
rewritten by means of non-linear map under several conditions. 
Mathematical treatment of the map enables us to discuss the condition on which the rankings of 
arbitrary two companies are reversed during the dynamics. 
The so-called `mismatch' between students and companies is discussed from both empirical and theoretical viewpoints. 
}

\section{Introduction}
\label{sec:1}
Deterioration of the employment rate is now one of the most serious problems in Japan 
\cite{keizaisanngyou,kouseiroudou,works}and various attempts to overcome these difficulties have been done by central or local governments. 
Especially, in recent Japan, the employment rate in young generation such as university graduates is getting worse.  
To consider the effective policy and to carry out it for sweeping away the unemployment uncertainty, it seems that we should simulate artificial labor markets in computers to reveal the essential features of the problem. 
In fact, in macroeconomics (labor science), there exist a lot of effective attempts to discuss the macroscopic properties \cite{Aoki,Boeri,Roberto,Fagiolo,Casares,Neugart} 
including so-called search theory \cite{Lippman,Diamond,Pissarides1985,Pissarides,Search}.
However, apparently, the macroscopic approaches lack of their microscopic viewpoint, namely, in their arguments, the behaviour of microscopic agents such as job seekers or companies are neglected. 

Taking this fact in mind, in our preliminary studies \cite{Chen2,Chen}, we proposed a simple probabilistic model based on the concept of statistical mechanics for stochastic labor markets, in particular, Japanese labor markets for university graduates. 
In these papers \cite{Chen2, Chen}, we showed that  a phase transition takes place in macroscopic quantities such as unemployment rate as the degree of high ranking preferential factor increases. 
These results are obtained at the equilibrium state of the labor market, however, the dynamical aspect seems to be important to reveal the matching process between the students and companies. 
Hence, in this paper, we shall focus on the dynamical aspect of our probabilistic labor market. 

This paper is organized as follows. 
In section \ref{sec:2}, we introduce our probabilistic model according to the references \cite{Chen2,Chen}.  
In the next section \ref{sec:3}, we show the aggregation probability of each company is described by a non-linear map. Using the knowledge obtained from the non-linear map, we discuss the condition on which the ranking of arbitrary two companies is reversed in successive two business years in section \ref{sec:4}, 
In section \ref{sec:5}, we discuss the global mismatch measurement, namely, ratio of job supply. 
We compare the result with the empirical evidence in recent Japan. 
In section \ref{sec:6}, we introduce a simple procedure to derive the analytic form of the aggregation probability at the steady state by means of `high-temperature expansion'. 
The last section \ref{sec:7} is summary.
\section{Model systems}
\label{sec:2}
According to our preliminary studies \cite{Chen2,Chen}, 
we shall assume that the following four 
points (i)-(iv) should be taken into account to 
construct the our labor markets. 
\begin{enumerate}
\item[(i)]
Each company recruits constant numbers of 
students in each business year. 
\item[(ii)] 
If the company 
takes too much or too less applications which are far beyond or 
far below the quota, the ability of the company to gather students in the next business year decreases. 
\item[(iii)]
Each company is apparently ranked according to 
various perspectives. 
The ranking information is available for all students.  
\item[(iv)]
A diversity of making decision of students 
should be taken into account by means of 
maximization of Shannon's entropy under some constraints. 
\end{enumerate}
To construct 
labor markets by considering the above four essential points, 
let us 
define the total 
number of companies as $K$ 
and each of them is distinguished 
by the label: $k=1,2, \cdots,K$. 
Then, the number of 
the quota of the company $k$ is 
specified by $v_{k}^{*}$. 
In this paper, we shall  
fix the value $v_{k}^{*}$ and regard the quota as 
a `time-independent' variable. 
Hence, the total quota (total job vacancy in society) in each 
business year $V$ is now given by 
\begin{equation}
V  =  \sum_{k=1}^{K}v_{k}^{*}.
\end{equation}
When we define 
the number of students by $N$ 
(each of the students is explicitly specified by the index $i$ as 
$i=1,2,\cdots, N$), one assumes that the $V$ 
is proportional to $N$ as 
$V  =  \alpha N$, where $\alpha$ stands for 
{\it job offer ratio} and 
is independent of $V$ and $N$. 
Apparently, 
for $\alpha  =V/N >1$, that is 
$V>N$, the labor market behaves as a `seller's market', 
whereas for $\alpha < 1$, 
the market becomes a `buyer's market'. 

We next define a sort of `energy function'  for each 
company which represents the ability (power) of 
gathering applicants in each business year $t$. 
The energy function is a nice bridge to link the labor market to physics. 
We shall first define the {\it local mismatch measurement}: $h_{k}(t)$ for 
each company  
$k$ $(k=1,2,\cdots,K)$ as 
\begin{equation}
h_{k}(t) =  
\frac{1}{V} 
|v_{k}^{*}-v_{k}(t)| = 
\frac{1}{\alpha N} 
|v_{k}^{*}-v_{k}(t)|,
\end{equation}
where  
$v_{k}(t)$ 
denotes the number of students who seek for the position in 
the company $k$ at the business year $t$ 
(they will post their own `entry sheet (CV)' to the company $k$). 
Hence, 
the local mismatch measurement $h_{k}(t)$ 
is the difference between 
the number of applicants $v_{k}(t)$ and the quota $v_{k}^{*}$. 
We should keep in mind that 
from the fact (i) mentioned before, 
the $v_{k}^{*}$ is 
a business year $t$-independent constant.  

On the other hand, we define the ranking of the company $k$ by $\epsilon_{k} (>1)$ 
which is independent of the business year $t$. 
Here we assume that 
the ranking of the company $k$ is higher 
if the value of $\epsilon_{k}$ is larger. 
In this paper, we simply set the value as  
\begin{equation}
\epsilon_{k} = 1+\frac{k}{K}.
\end{equation}
Namely, 
the company $k=K$ is the highest ranking 
company and 
the company $k=1$ is the lowest. 
Then, 
we define the energy function of our labor market for each company $k$ as 
\begin{equation}
E(\epsilon_{k}, h_{k}; t)  \equiv  
-\gamma \log \epsilon_{k} + 
\sum_{l=1}^{\tau}
\beta_{l}h_{k}(t-l).
\label{eq:energy}
\end{equation}
From the first term appearing in the 
right hand side of the above energy function (\ref{eq:energy}), 
students tend to apply their entry sheets 
to relatively high ranking companies. 
However, the second term in (\ref{eq:energy}) 
acts as the `negative feedback' 
on the first ranking preference to decrease the probability at the next business year $t+1$ for  
the relatively high ranking company gathering the applicants at the previous business year $t$.  
Thus, the second term is actually regarded as a negative feedback on 
the first term. The ratio $\gamma/\beta_{l},\,(l=1,\cdots,\tau)$ 
determines to what extent 
students take into account the history of 
the labor market. 
In this paper, we simply set $\beta_{1}=\beta, 
\beta_{2}=\cdots=\beta_{\tau}=0$, 
namely, we assume that each student consider the latest  result in 
the market history. 

We next adopt the assumption (iv). 
In order to quantify the diversity of 
making decision 
by students, we introduce the following 
Shannon's entropy: 
\begin{equation}
H = 
-\sum_{k=1}^{K}
P_{k}(t) \log P_{k}(t)
\end{equation}
Then, let us consider the probability $P_{k}(t)$ that 
maximizes the above $H$ under the normalization 
constraint $\sum_{k=1}^{K}P_{k}(t)=1$. 
To find such $P_{k}(t)$, we maximize the functional $f$ 
with Lagrange multiplier $\lambda$, 
$f = 
-\sum_{k=1}^{K}
P_{k}(t) \log P_{k}(t)+\lambda
\{
\sum_{k=1}^{K}P_{k}(t)-1\}$, 
with respect to $P_{k}(t)$. 
A simple algebra gives the solution 
\begin{equation}
P_{k}(t) = 
\frac{1}{K}. 
\end{equation}
This implies that 
the most diverse 
making decision by students 
is realized by a random selection from among 
$K$ companies with probability $1/K$. 
(It should be noted that we set $K {\rm e}^{\lambda-1} =1$).

On the analogy of the Boltzmann-Gibbs distribution in statistical mechanics, 
we add an extra constraint in such a way that 
the expectation of energy function over the 
probability $P_{k}(t)$ 
is constant for each business year $t$, 
namely,  $E = \sum_{k=1}^{K}P_{k}(t)E(\epsilon_{k},h_{k};t)$. 
Taking into account this constraint by means of 
another Lagrange multiplier $\lambda^{'}$ and 
maximizing the functional 
\begin{eqnarray}
f & = &  
-\sum_{k=1}^{K}
P_{k}(t) \log P_{k}(t)+\lambda
\left\{
\sum_{k=1}^{K}P_{k}(t)-1
\right\} +
\lambda^{'}
\left\{
E = \sum_{k=1}^{K}P_{k}(t)E(\epsilon_{k},h_{k};t)
\right\}
\end{eqnarray}
with respect to 
$P_{k}(t)$, we have the probability 
$P_{k}(t)$ 
that the company $k$ gathers their applicants   
at time $t$ as 
\begin{equation}
P_{k}(t)   =  
\frac{
{\exp}
\left[
-E(\epsilon_{k}, h_{k}(t-1))
\right]}{Z}, \,\,E(\epsilon_{k},h_{k}(t-1)) \equiv  -\gamma \log \epsilon_{k}+\beta h_{k}(t-1)
\label{eq:prob} 
\end{equation}
where we defined 
$Z \equiv 
\sum_{k=1}^{K}
{\exp}[-E(\epsilon_{k}, h_{k}(t-1))]$ stands for the normalization constant for 
the probability. 
The parameters $\gamma$ and $\beta$  
specify the probability from the 
macroscopic point of view. 
Namely, 
the company $k$ having relatively small 
$h_{k}(t)$ 
can gather a lot of applicants 
in the next business year and 
the ability is controlled by 
$\beta$ (we used the assumption (ii)). 
On the other hand, 
the high ranked company can gather 
lots of applicants and 
the degree of the ability is specified by 
$\gamma$  (we used the assumption (iii)). 

We should notice that 
for the probability $P_{k}(t)$, 
each student $i$ 
decides to post their entry sheet to 
the company $k$ at time $t$ as 
\begin{eqnarray}
a_{ik}(t) & = & 
\left\{
\begin{array}{ll}
1 & (\mbox{with}\,\, P_{k}(t))\\
0 & (\mbox{with}\,\, 1-P_{k}(t)) 
\label{eq:aik}
\end{array}
\right.
\end{eqnarray}
where $a_{ik}(t)=1$ means 
that the student $i$ posts his/her 
entry sheet to the company $k$ and 
$a_{ik}(t)=0$ denotes that 
he/she does not. 

We might consider the simplest case in which 
each student $i$ posts their single entry sheet to 
a company with probability $P_{k}(t)$.
Namely, $P_{k}(t)$ is independent of $i$. 
From the definition, 
the company $k$ gets the entry sheets 
$N P_{k}(t)$ on average and 
the total number of 
entry sheets is 
$N\sum_{k=1}^{K}P_{k}(t)=N$. 
This means that each student  
applies only once on average to one of the $K$ companies. 
We can easily extend this situation by assuming that 
each student posts his/her entry sheets $a$-times on average. 
In this sense, the situation changes in such a way that 
the company $k$ takes 
$aNP_{k}(t)$-entry sheets on average. 

Now it is possible for us to evaluate how many acceptances 
are obtained by a student and 
let us define the number by $s_{i}(t)$ for 
each student  $i=1,\cdots,N$. 
Then, we should notice that 
the number of 
acceptances for the student $i$ is 
defined by 
$s_{i}(t) = \sum_{k=1}^{K}s_{ik}(t)$ with 
\begin{equation}
P(s_{ik}(t)=1|a_{ik}(t)) = 
\Theta(v_{k}^{*}-v_{k}(t)) \delta_{a_{ik}(t),1} +  
\frac{v_{k}^{*}}{v_{k}^{*}(t)}\Theta (v_{k}(t)-v_{k}^{*}) \delta_{a_{ik}(t),1}
\label{eq:def_s} 
\end{equation}
 and $P(s_{ik}(t)=0|a_{ik}(t)) =1-P(s_{ik}(t)=1|a_{ik}(t))$, 
 where 
 $\Theta (\cdots)$ denotes the 
 conventional step function and 
 one should remember that 
 we defined new variables $a_{ik}(t)$ 
 which give $a_{ik}(t)=1$ when 
 the labor $i$ posts the entry sheet to 
 the company $k$ and $a_{ik}(t)=0$ vice versa.  
 Thus, the first term of (\ref{eq:def_s}) means that 
 the $s_{ik}(t)$ takes $1$ with unit probability when 
 the labor $i$ posts the sheet to the company $k$ and 
 the total number of sheets gathered by the company $k$ 
 does not exceed the quota $v_{k}^{*}$. 
 On the other hand, the second term 
 means that the $s_{ik}(t)$ takes $1$ 
 with probability $v_{k}^{*}/v_{k}(t)$ even if 
 $v_{k}(t)>v_{k}^{*}$ holds. 
 In other words, 
 for $v_{k}(t)>v_{k}^{*}$, 
 the informally accepted $v_{k}^{*}$ students are randomly selected from 
 $v_{k}(t)$ candidates.

The probability (\ref{eq:prob}) describes the microscopic 
foundation of our labor markets. 
In the next section, we show that the update rule 
of the probability $P_{k}(t)$ is regarded as a non-linear map 
in the limit of $N \to \infty$. 
\section{Non-linear map for the aggregation probability}
\label{sec:3}
In this section, we derive a non-linear map for the aggregation probability 
$P_{k}(t)$. 
Under the simple assumption 
$v_{k}^{*}(=\mbox{const.})$, 
we have 
 $V=\sum_{k=1}^{K}v_{k}^{*}=Kv_{k}^{*}$. 
 On the other hand, from the definition of 
 job offer ratio $V=\alpha N$, 
 $\beta v_{k}^{*}/V=\beta/K$ and  
 $(\beta/V)v_{k}(t-1) = 
 (\beta/\alpha)
 (v_{k}(t-1)/N) = 
 (\beta/\alpha)P_{k}(t-1)$ should be satisfied. 
 When $NP_{k}, NP_{k}(1-P_{k})$ are large enough,  
the binomial distribution of $v_{k}(t)$, that is, 
$P(v_{k})={}_{N}C_{v_{k}}P_{k}^{v_{k}}(1-P_{k})^{N-v_{k}}$ 
could be approximated by a normal distribution 
with mean $NP_{k}$ and variance $NP_{k}(1-P_{k})$. 
Namely, 
\begin{equation}
v_{k}(t) = 
NP_{k} + \sqrt{NP_{k}(1-P_{k})}{\cal N}(0,1).
\end{equation}
This reads $v_{k}(t)/N = 
P_{k} + \sqrt{P_{k}(1-P_{k})/N}
{\cal N}(0,1)$ and 
the second term can be dropped 
in the limit $N \to \infty$ as 
$\lim_{N \to \infty} v_{k}(t)/N =P_{k}(t)$.
 Thus  the probability 
 $P_{k}(t)$ is now rewritten by  
 \begin{equation}
 P_{k}(t) = 
 \frac{{\exp}
 \left[
 \gamma \log 
 \left(
 1+\frac{k}{K}
 \right)-
 \frac{\beta}{\alpha}
 |P_{k}(t-1)-\frac{\alpha}{K}|
 \right]}
 {
 \sum_{k=1}^{K}
 {\exp}
 \left[
 \gamma \log 
 \left(
 1+\frac{k}{K}
 \right)-
 \frac{\beta}{\alpha}
 |P_{k}(t-1)-\frac{\alpha}{K}|]
 \right]}.
 \label{eq:chen4}
 \end{equation}
 This is nothing but a non-linear map for the 
 probability $P_{k}(t)$. 
 Hence, 
 one can evaluate the time-evolution of 
 the aggregation probability $P_{k}(0) \to P_{k}(1) \to \cdots \to P_{k}(t) \to \cdots$ 
 by solving the map (\ref{eq:chen4}) recursively from an arbitrary 
 initial condition, say,  $P_{1}(0)=P_{2}(0)=\cdots =P_{K}(0)=1/K$.
 \begin{figure}[ht]
 \begin{center}
 \includegraphics[width=5.8cm]{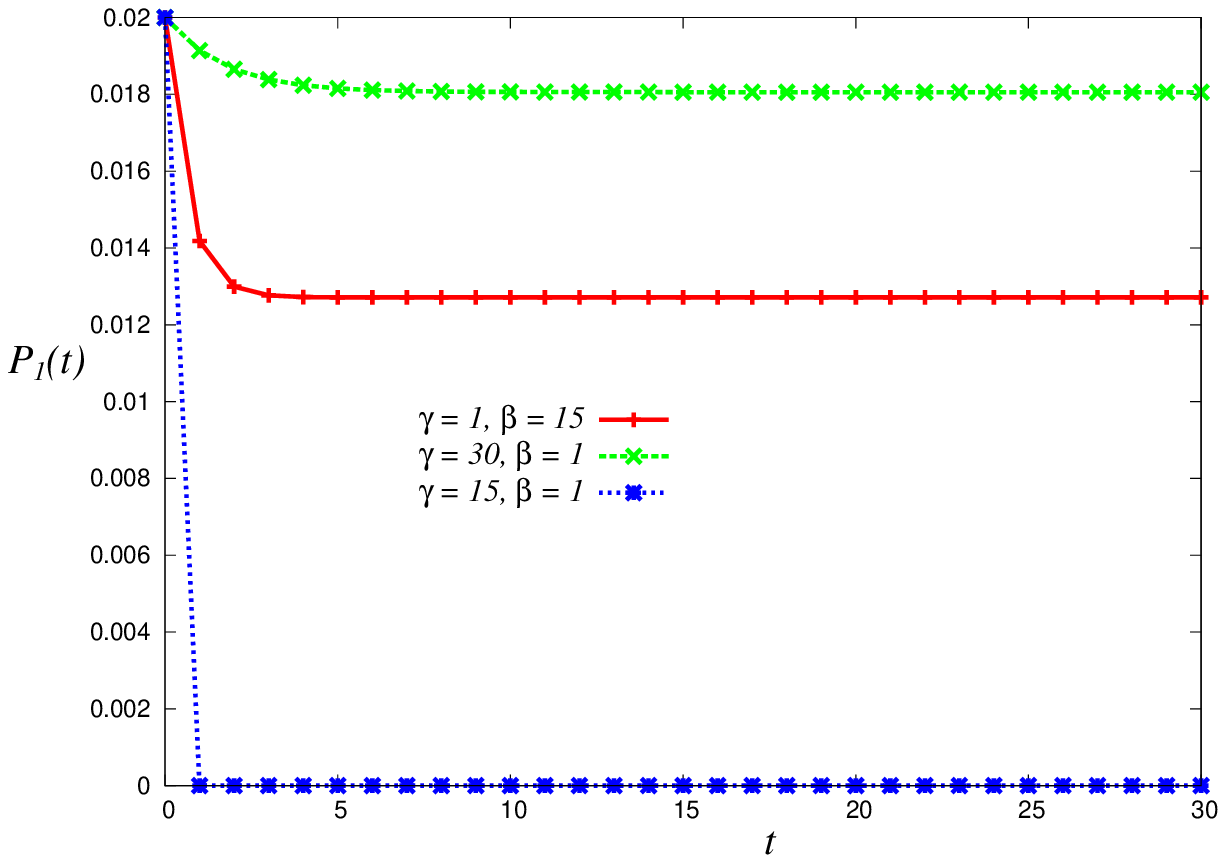}
 \includegraphics[width=5.8cm]{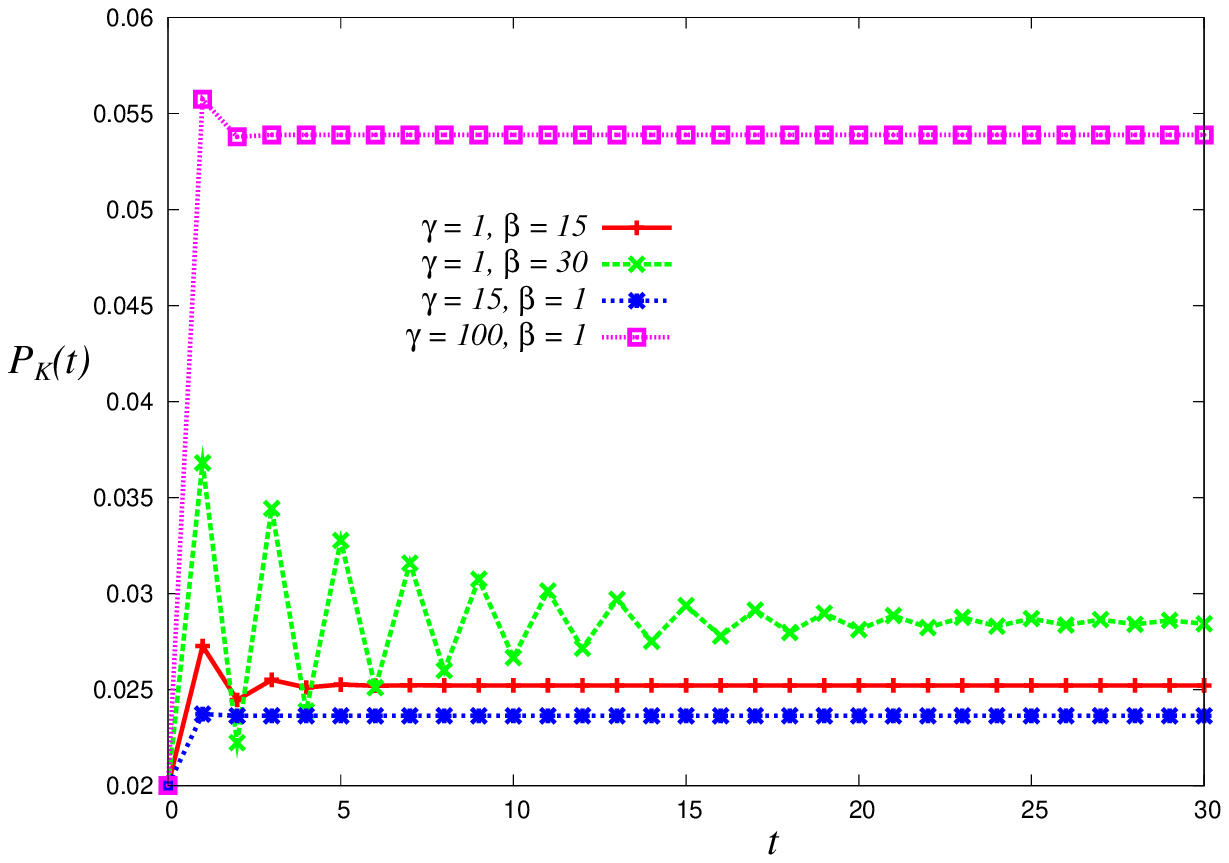}
 \end{center}
 \caption{\footnotesize 
 The time evolution of probability for the lowest ranking company 
 $P_{1}(t)$ (left) and the highest ranking company $P_{K}(t)$ (right). 
 We set $K=50, \alpha=1$ and varied the values of parameters $\beta,\gamma$. 
 }
 \label{fig:fg1}
 \end{figure}
 In Fig. \ref{fig:fg1}, 
 we plot the time evolution of the aggregation probability for the lowest ranking company 
 $P_{1}(t)$ (left) and the highest ranking company $P_{K}(t)$ (right). 
 From this figure, we easily find that the $P_{K}(t)$ 
 oscillates for $\beta \gg \gamma$ due to the second term of the energy function (\ref{eq:prob}), 
 namely, the negative feed back acts on the high-ranking preference term $-\gamma \log \epsilon_{k}$.  
 
 When we define the state satisfying $P_{1}(t) < \epsilon \equiv 10^{-5}$
 as  a kind of `business failures', 
 the failure does not take place until  $\gamma \simeq 10.3 \equiv \gamma_{\rm c}$. 
 However, the failure emerges the parameter $\gamma$ 
 reaches the critical value $\gamma=\gamma_{\rm c}$. 
 \section{Ranking frozen line}
 \label{sec:4}
\begin{figure}[ht]
 \begin{center}
  \includegraphics[width=7cm]{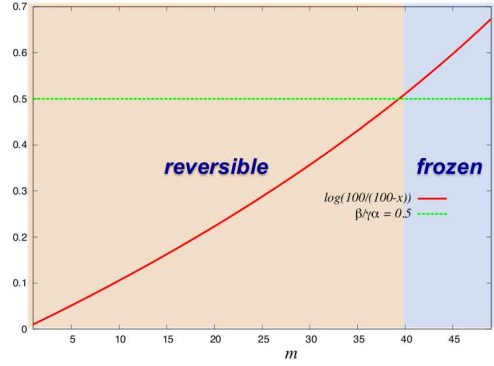}
 \end{center}
 \caption{\footnotesize 
 The boundary of frozen ranking. 
 The solid and the broken lines 
 are $\log (2K/(2K-m))$ and $\beta /\gamma \alpha$, respectively. 
We set $K=50, \beta /\gamma \alpha =0.5$. 
 }
 \label{fig:fg2}
 \end{figure}

 Here we discuss 
 the condition on which 
the order of probabilities $P_{K}(t) > P_{K-m}(t)$ is reversed 
at time $t+1$ as $P_{K}(1+1) < P_{K-m}(t+1)$. 
After simple algebra, we find the condition explicitly as 
\begin{equation}
\log 
\left(
\frac{2K}{2K-m}
\right) > \frac{\beta}{\gamma \alpha}.
\label{eq:condition1}
\end{equation}
From this condition, we are confirmed that if the strength of market history 
$\beta$ is strong, the condition is satisfied easily, whereas if 
the ranking factor $\gamma$ or job offer ratio $\alpha$ is large, 
the condition is hard to be satisfied, namely, the ranking 
of the highest and $m$-th highest is `frozen'. 
In Fig. \ref{fig:fg2}, we draw the example. 
We set $K=50, \beta/\alpha \gamma=0.5$. 
From this figure, we actually find, for example, 
the highest ranking company ($m=0$) and 
$45$-th highest ranking company ($m=45$) cannot be reversed. 
 \section{Global mismatch measurement}
 \label{sec:5}
We next discuss the global mismatch measurement between students and companies. 
For this purpose, we should evaluate ratio of job supply defined as 
\begin{equation}
\Omega  \equiv \frac{1}{V}\sum_{k=1}^{K}(v_{k}^{*}-m_{k})
=1-\frac{1}{V}\sum_{k=1}^{K}m_{k}.
\label{eq:Omega_alpha}
\end{equation}
Here we assumed that the student who gets multiple informal acceptances chooses 
the highest ranking company to go, namely, 
$\tilde{l}_{i}={\rm argmax}_{l} \epsilon_{l}\delta_{s_{il},1}$, where the label $s_{il}$ takes $1$ 
when the student $i$ obtains the informal acceptance from the company $l$ and it takes $0$ if not. 
Then, the new employees of company $k$, namely, 
the number of students whom the company $k$ obtains is given by 
$m_{k}=\sum_{i=1}^{N}s_{ik}\delta_{k,\tilde{l}_{i}}$.

On the other hand, it should be noticed that 
the number of all newcomers (all new employees) in the society is 
given by 
\begin{equation}
\sum_{k=1}^{K}
m_{k} = N-N \sum_{i=1}^{N}
\delta_{s_{i},0}=N(1-U).
\label{eq:U_alpha1}
\end{equation}
From (\ref{eq:Omega_alpha}) and (\ref{eq:U_alpha1}), 
we have the linear relationship between 
the unemployment rate $U$ and the 
ratio of job supply $\Omega$ as 
\begin{equation}
U = 
\alpha \Omega + 
1-\alpha. 
\label{eq:U_alpha}
\end{equation}
It should be noted that the location of  
a single realization 
$(U, \Omega)$ 
is dependent on 
$\beta,\gamma$ and $\alpha$ 
through the $\Omega$.

Large mismatch is specified by the area in which both $U$ and $\Omega$ are large.  
We plot the empirical evidence and our result with $\gamma=\beta=1$ in Fig. \ref{fig:fg3}. 
 \begin{figure}[ht]
 \begin{center}
 \includegraphics[width=5.2cm]{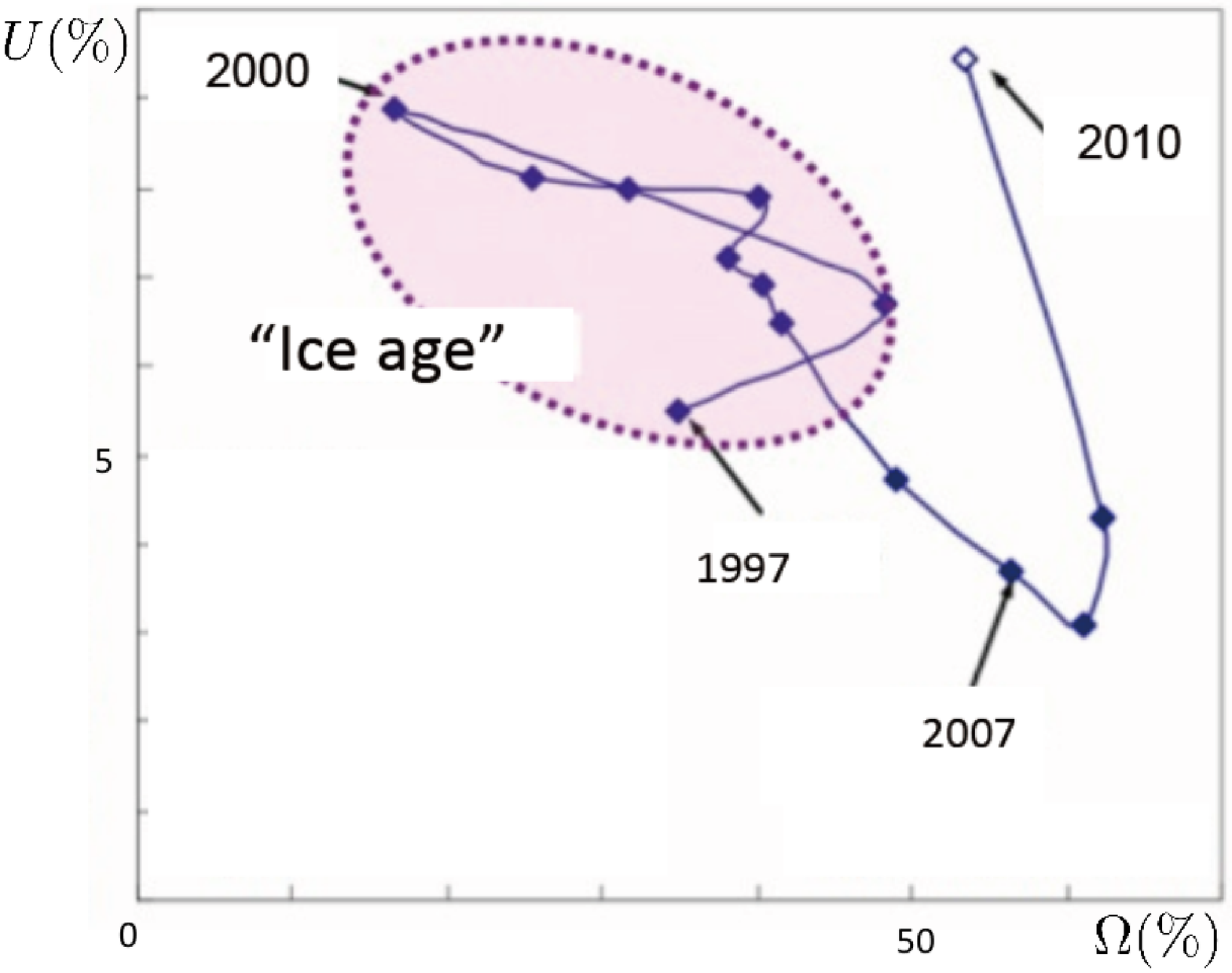}
 \includegraphics[width=5.8cm]{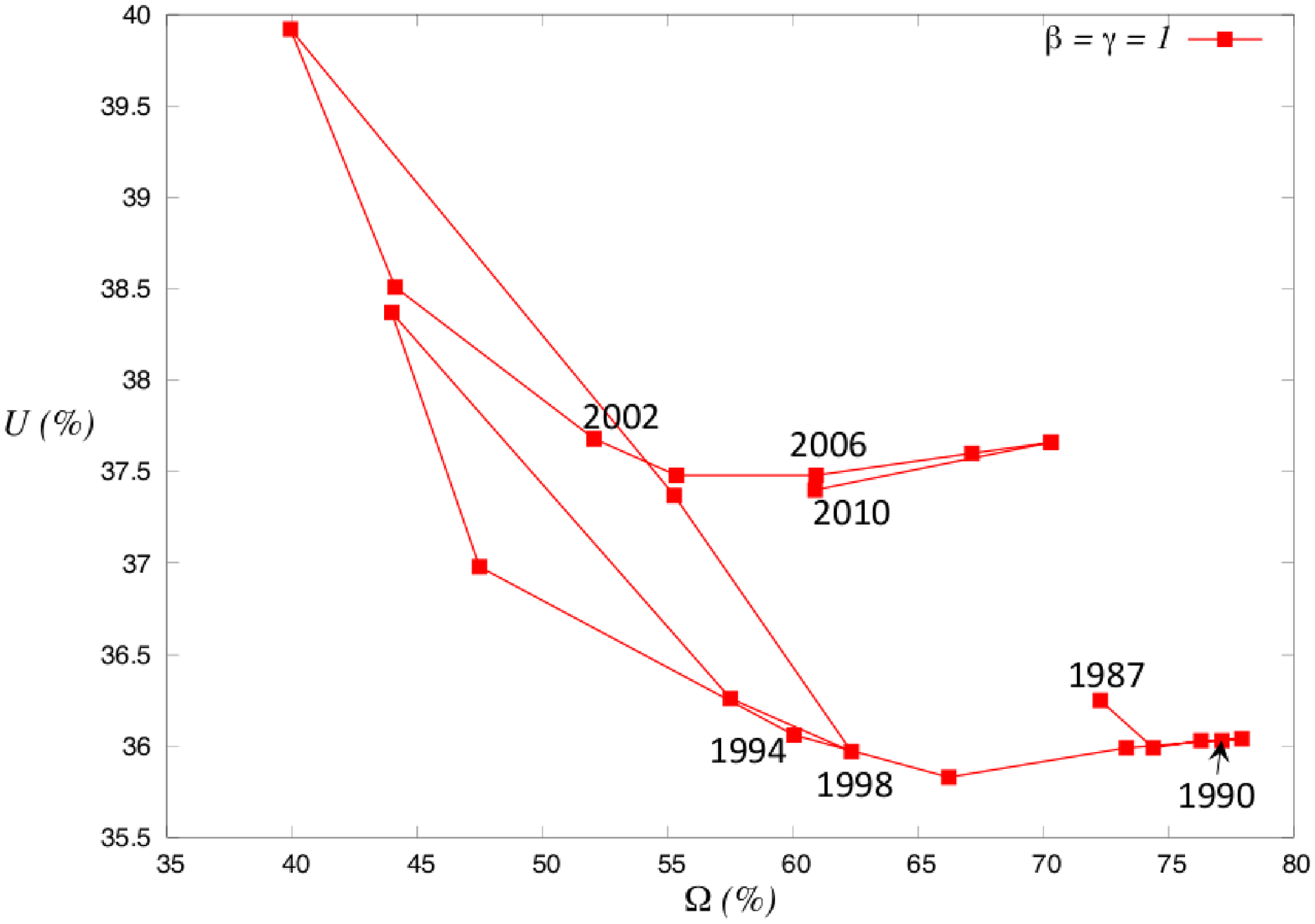}
 \end{center}
 \caption{\footnotesize 
 The relationship between 
 $U$ and $\Omega$ for the past Japanese labor market for university graduates. 
 The left panel is empirical evidence provided by `Japan-Insight' by Mizuho Research Institute (2011). The right panel is obtained by our toy model. 
 We used the empirical data for the job offer ratio $\alpha$ and set the `unobservable' parameters as $\beta=\gamma=1$ (equally weights).
 }
 \label{fig:fg3}
 \end{figure}
We used the empirical data for the job offer ratio $\alpha$ and we simply set $\beta=\gamma=1$. 
 We find that qualitative behaviours for both empirical and theoretical are similar but quantitatively they are different. 
 This is because we cannot access the information about macroscopic variables $\beta,\gamma$ 
 and the result might depend on the choice. 
 Estimation for these variables by using empirical data should be addressed as our future problem.
\section{Aggregation probability at `high temperature'}
\label{sec:6}
In the previous sections, we investigate the 
dynamics of $P_{k}(t)$ as a non-linear map. 
We carried out numerical calculations 
to evaluate the steady state. 
Here we attempt to 
derive the analytic form of the aggregation probability 
at the steady state by means of 
{\it high-temperature expansion} 
for the case of $\gamma,  \beta/\alpha \ll 1$. 
\subsection{The high temperature expansion}
Let us first consider the zero-th approximation. 
Namely, we shall rewrite the aggregation probability 
in terms of ${\rm e}^{x} = 1$.  
This immediately reads $Z=\sum_{k=1}^{K}1=K$ and 
we have 
\begin{equation}
P_{k}^{(0)}=\frac{1}{K}.
\end{equation}
This is nothing but 
`random selection' by students in the high temperature limit $\gamma, \beta/\alpha \to 0$.  
This is a rather trivial result and this type of procedure should be proceeded until 
we obtain non-trivial results. 

In order to proceed to carry out our approximation 
up to the next order $P_{k}^{(1)}$, 
we first consider the case of 
$P_{k}^{(1)}>\alpha/K$. 
By making use of ${\rm e}^{x}=1+x$, 
we obtain $Z =  
K + 
\gamma 
\sum_{k=1}^{K}
\log (1+k/K) -(\beta/\alpha) +\beta$. 
Hence, if one notices that 
\begin{equation}
\sum_{k=1}^{K}
\log 
(1+k/K)  =  
K \cdot \frac{1}{K}
\sum_{k=1}^{K}
\log (1+k/K) \simeq    
K \int_{0}^{1} \log (1+x)dx =  
K(2\log 2-1)
\end{equation}
holds for for $K \gg 1$, 
the normalization constant 
$Z$ leads to $Z = 
K(1+2\gamma \log 2-\gamma)
-(\beta/\alpha)+\beta$. 
By setting $P_{k}(t)=P_{k}(t-1)=P_{k}^{(1)}$ (steady state), 
we have 
\begin{equation}
P_{k}^{(1)} = 
\frac{1+\gamma \log (1+k/K) -\frac{\beta}{\alpha}P_{k}^{(1)}+\frac{\beta}{K}}
{Z}. 
\end{equation}
By solving the above equation 
with respect to $P_{k}^{(1)}$, 
one finally has 
\begin{equation}
P_{k}^{(1)} = 
\frac{1+\frac{\beta}{K}+\gamma \log (1+k/K)}
{K(1+2\gamma \log 2 -\gamma)+\beta}. 
\end{equation}
For the case of $P_{k}^{(1)}<\alpha/K$, 
the same argument as the above gives 
\begin{equation}
\hat{P}_{k}^{(1)} \equiv  
\frac{1-\frac{\beta}{K}+\gamma \log (1+k/K)}
{K(1+2\gamma \log 2 -\gamma)-\beta}. 
\end{equation}
It should be noted that 
$P_{k}^{(1)}$ is independent of 
the job offer ratio $\alpha$. 
We also should notice that  
$P_{k}^{(1)} = \hat{P}_{k}^{(1)}=1/K=P_{k}^{(0)}$ 
is recovered by setting $\beta=0$ in the $P_{k}^{(1)}$. 
\begin{figure}[ht]
\begin{center}
\includegraphics[width=5.8cm]{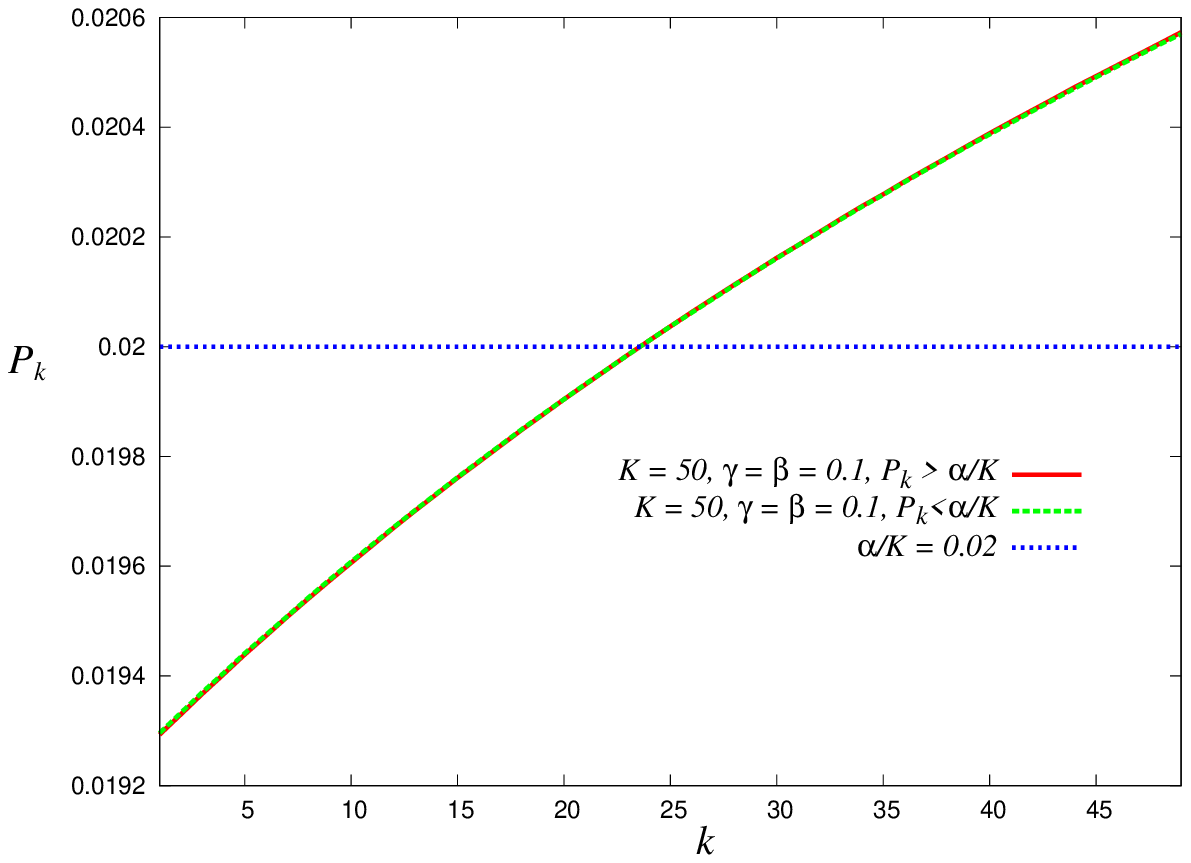}
\includegraphics[width=5.8cm]{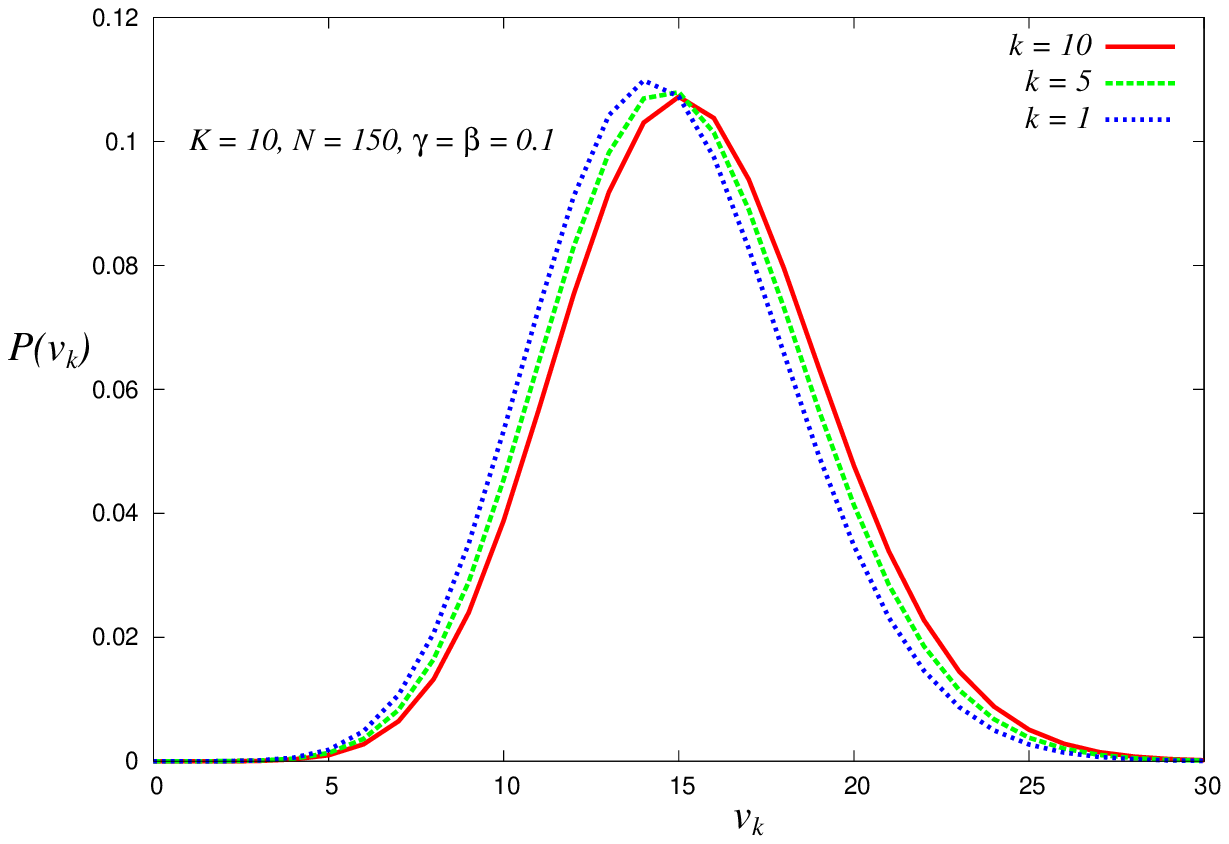}
\end{center}
\caption{\footnotesize 
The first-order approximation 
$P_{k}^{(1)}=P_{k}$ at high temperature (left). 
The right panel shows the corresponding probability 
$P(v_{k})$.  
 }
\label{fig:fg3}
\end{figure}
We show the result in Fig. \ref{fig:fg3}.

We can proceed to carry out the above approximation systematically 
by using the Tayler's expansion of 
exponential ${\exp}(x)$. 
However, unfortunately, when we try to 
obtain the second order approximation, 
we encounter the terms such as 
$\sum_{k} P_{k} \log (1+k/K)$. 
Apparently, it is impossible for us to evaluate this term 
until we obtain $P_{k}$. 
Hence, we shall replace $P_{k}$ in the above expression 
by the first order solution $P_{k}^{(1)}$. 
We also replace 
$(P_{k})^{2}$ by $(P_{k}^{(1)})^{2}$, 
and by using the fact 
\begin{equation}
\frac{1}{K} \sum_{k=1}^{K} \{\log (1+k/K)\}^{2} \simeq 
\int_{0}^{1}\{\log (1+x)\}^{2}dx= 
2\{(\log 2)^{2}-2\log 2 +1\}
\end{equation}
we obtain $P_{k}^{(2)}$ as 
\begin{equation}
P_{k(\pm)}^{(2)}=
\frac{\psi_{\pm} + 
\frac{\gamma^{2}}{2}\{\log (1+\frac{k}{K})\}^{2}+\frac{\beta^{2}}{2K^{2}}
\{1+ (\frac{K +\beta}{\alpha B})^{2}\}}
{\frac{\beta (\psi_{\pm} \mp 1)}{\alpha} + B 
\gamma^{2}K\chi_{1} - 
\frac{\beta \gamma}{\alpha B}
(2K\gamma \chi_{1}+(K + \beta)\chi_{2})}
\end{equation}
where we defined 
\begin{eqnarray}
\chi_{1} & \equiv  & (\log 2)^{2}-2\log 2 +1, \,\,
\chi_{2} \equiv 2\log 2-1 \\
\psi_{\pm} & \equiv & \gamma \log (1+k/K)  \pm (\beta/K) +1, \,\,
B \equiv K\{2\gamma \log 2 -\gamma + 1\}+\beta.
\end{eqnarray}
We also defined 
$P_{k+}^{(2)}$ as a solution for 
$P_{k}>\alpha/K$, 
whereas 
$P_{k-}^{(2)}$ is a solution for $P_{k}<\alpha/K$. 

 As we saw in the above argument of high-temperature expansion,  
 one can obtain the higher-order approximation 
 for the aggregation probability $P_{k}$ systematically. 
 However, we should bear in mind that 
 the above approximation should be broken down 
 for $\gamma, \beta/\alpha \simeq 1$. 
\subsection{Analytic solution for unemployment rate}
We next show that 
one can obtain the analytic solution for the employment rate 
in terms of the aggregation probability at high temperature. 
Here we show the solution for the first order approximation 
$P_{k}^{(1)}$ of the aggregation probability, however, 
the extension to the higher-order is straightforward. 

As we already saw, for $P_{k}=P_{k}^{(1)}$ and 
 $P_{k}>\alpha/K$,  
$NP_{k}, NP_{k}(1-P_{k}) \gg 1$, we have 
$\lim_{N \to \infty} v_{k}/N =P_{k}$.
Namely, 
the probability that 
the company $k$ obtains $v_{k}$-entry sheets 
is written as 
$P(v_{k}) =  
\delta (v_{k}-NP_{k})\Theta (P_{k}-\alpha/K)  + 
\delta (v_{k}-N\hat{P}_{k})
\Theta (\alpha/K-P_{k})$. 
We should notice that 
there exists a point $k$ at which 
$P_{k}=\hat{P}_{k}=\alpha/K$ holds. 

When we define $P(s_{ik}=1|v_{k})$ 
as the probability that 
student $i$ receives an informal acceptance 
from the company $k$ which gathered  $v_{k}$-entry sheets 
from the students including the student $i$, 
the probability $P(s_{ik}=1|v_{k})$ is given by
\begin{eqnarray}
P(s_{ik}=1|v_{k}) & = & 
P(s_{ik}=1|a_{ik}=1,v_{k}>v_{k}^{*}) 
P(a_{ik}=1) 
\Theta (v_{k}-v_{k}^{*})  \nonumber \\
\mbox{} & + &  
P(s_{ik}=1|a_{ik}=1,v_{k}<v_{k}^{*}) P(a_{ik}=1) 
\Theta (v_{k}^{*}-v_{k}).
\end{eqnarray}
Then, taking into account the fact 
$P(s_{ik}=1|a_{ik}=1,v_{k}>v_{k}^{*})=v_{k}^{*}/v_{k}, 
P(s_{ik}=1|a_{ik}=1,v_{k}<v_{k}^{*})=1$, 
and  $P(a_{ik}=1)=P_{k}$, we immediately obtain 
\begin{equation}
P(s_{ik}=1|v_{k}) = 
\left(
\frac{v_{k}^{*}}
{v_{k}}
\right) P_{k} \Theta (v_{k}-v_{k}^{*}) +  
\hat{P}_{k} \Theta (v_{k}^{*}-v_{k}). 
\end{equation}
Obviously, the above 
$P(s_{ik}=1|v_{k})$ depends on $v_{k}$. 
Hence, we calculate 
the average of $P(s_{ik}=1|v_{k})$ over the probability $P(v_{k})$, 
namely 
$P(s_{ik}=1) =  
\sum_{v_{k}}
P(v_{k})
P(s_{ik}=1|v_{k})$. 
Then, we have 
\begin{equation}
\phi_{k} \equiv 
\frac{\alpha}{K}
\Theta (
P_{k}-\alpha/K) + 
\hat{P}_{k}\Theta (\alpha/K-\hat{P}_{k})
\end{equation}
\begin{figure}[ht]
\begin{center}
\includegraphics[width=7cm]{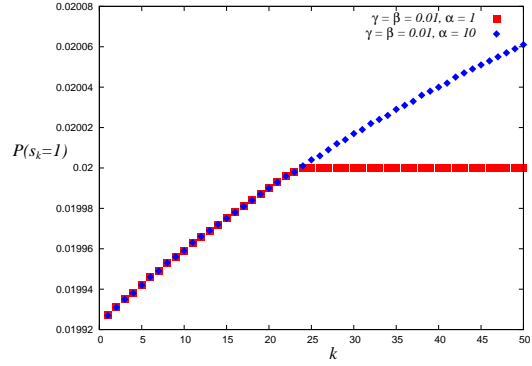}
\end{center}
\caption{\footnotesize 
$P(s_{k}=1) \equiv \phi_{k}$ evaluated by means of 
$P_{k}^{(1)}$.}
\label{fig:fg9}
\end{figure}
where we canceled the index $i$ because 
it is no longer dependent on specific student $i$ and 
we define $\phi_{k}$ as the probability that 
an arbitrary student gets the informal acceptance from the company 
$k$. 
In Fig. \ref{fig:fg9}, 
we plot the typical behaviour of 
the $\phi_{k}$. 

Therefore, 
the probability 
that an arbitrary student who sent 
three entry sheets to three companies $k, l,m$ 
could not get any informal acceptance is 
given by 
$(1-\phi_{k})(1-\phi_{l})(1-\phi_{m})/3!$.  
From this observation, 
we easily conclude that 
analytic form of the unemployment rate is 
obtained as 
\begin{equation}
U =  
\sum_{k_{1}<k_{2}<\cdots<k_{a}}^{K}
\frac{(1-\phi_{k_{1}}) \cdots (1-\phi_{k_{a}})}
{K(K-1) \cdots (K-a+1)}. 
\end{equation}
%
 \section{Summary}
 \label{sec:7}
 In this paper, we introduced a toy probabilistic model to analyse job-matching processes in recent Japanese labor markets for 
university graduates by means of statistical physics. 
We succeeded in deriving a non-linear map with respect to 
the aggregation probability. 
By analyzing the map and evaluating the steady states by means of high-temperature expansion, 
we could discuss several analytic results and their mathematical properties. 
Global mismatch between students and companies was discussed from 
both empirical and theoretical viewpoints. 
However, as we saw in Fig. \ref{fig:fg3}, there still exists a gap between 
our result and the empirical evidence. 
To bridge the gap, 
we should estimate the non-observable parameters 
such as $\gamma, \beta$ using 
the data collected by appropriate questionnaires. It should be addressed as one of our future studies.  
\subsection*{Acknowledgement}
We thank Enrico Scalas and Giacomo Livan for valuable discussion. 
This work was financially supported by 
Grant-in-Aid for Scientific Research (C) of Japan Society for the Promotion of Science, No. 22500195. 

\end{document}